\documentclass[prb,twocolumn,showpacs,preprintnumbers,amsmath,amssymb]{revtex4}

\usepackage{graphicx}
\usepackage{dcolumn}
\usepackage{bm}

\begin{document}


\title{Multiple phase transitions in single-crystalline Na$_{1-\delta}$FeAs}

\author{G. F. Chen}
\author{W. Z. Hu}
\author{J. L. Luo}
\author{N. L. Wang}

\affiliation{Beijing National Laboratory for Condensed Matter
Physics, Institute of Physics, Chinese Academy of Sciences,
Beijing 100190, People¡¯s Republic of China}


\begin{abstract}
Specific heat, resistivity, susceptibility and Hall coefficient
measurements were performed on high-quality single crystalline
Na$_{1-\delta}$FeAs. This compound is found to undergo three
successive phase transitions at around 52, 41, and 23 K, which
correspond to structural, magnetic and superconducting
transitions, respectively. The Hall effect result indicates the
development of energy gap at low temperature due to the occurrence
of spin-density-wave instability. Our results provide direct
experimental evidence of the magnetic ordering in the nearly
stoichiometric NaFeAs.

\end{abstract}

\pacs{74.70.-b,74.25.Gz,74.25.Fy}


\maketitle

The recent discovery of superconductivity with transition
temperature T$_c\sim$  26 K in LaFeAsO$_{1-x}$F$_x$ (abbreviated
as 1111) has attracted a great deal of research
interest.\cite{Kamihara08} Substituting La with other rare earth
elements dramatically enhances the T$_c$ up to 41-55
K.\cite{Chen1, ChenXH, Ren1, Wang} At room temperature, all these
parent compounds crystallize in a tetragonal ZrCuSiAs-type
structure, which consists of alternate stacking of edge-sharing
Fe$_2$As$_2$ tetrahedral layers and La$_2$O$_2$ tetrahedral layers
along c-axis. Soon after this discovery, another group of
compounds AFe$_2$As$_2$(A=Ba, Sr, Ca)(122), which crystallize in a
tetragonal ThCr$_2$Si$_2$-type structure with identical
Fe$_2$As$_2$ tetrahedral layers as in LaFeAsO, were also found to
be superconducting with T$_c$ up to 38 K upon hole
doping.\cite{Rotter2, Chen, Canfield}

In these so called ``1111" and ``122" compounds, it is well known
that the Fe ions tend to form magnetically ordered state and
participate in building up a high density of states at the Fermi
level, which is responsible for the superconductivity.\cite{Dong}
However, very recently, LiFeAs and NaFeAs (111), with the PbFCl
structure type and containing Fe$_2$As$_2$ tetrahedral layers,
were reported to be superconducting (T$_c$ $\sim$ 18 K and
9$\sim$25 K, respectively) at ambient pressure without purposely
doping carriers, and no signature of magnetic order was detected
at temperature up to 300 K.\cite{Jin1, Zhu1, Zhu2, Parker} If the
absence of magnetic order is an intrinsic property for the ``111"
compounds, it would strongly challenge the generally accepted
picture that magnetic fluctuations play a crucial role in the
superconducting pairing. While band structure calculation
\cite{Jishi, Singh, Nekra} on stoichiometric LiFeAs/NaFeAs
suggests that they should be similar to the parent compounds of
``1111" and ``122", and they should display magnetic orders rather
than superconductivity in the ground state.

Experimentally, it is difficult to synthesize the stoichiometric
LiFeAs and NaFeAs, because of the evaporation loss of Li/Na during
the high-temperature reaction. It is reasonably convinced that
carriers are introduced in LiFeAs/NaFeAs by self-doping due to
Li/Na deficiencies. It has been confirmed in the Li deficient
sample Li$_{1-\delta}$FeAs.\cite{Jin1, Zhu1} The absence of
superconductivity was indeed found in the nearly stoichiometric
LiFeAs synthesized by high pressure method. However, there is
little experimental evidence for structural/magetic transition. It
may be due to Li deficiency, which introduces the level of carrier
density close to a ``critical value" just suppressing the
spin-density-wave (SDW) order.

It is of great interest to investigate the electronic properties
of the high quality single crystal of these materials (111) and
compare them with ``1111" and ``122" systems. Here we report on a
comprehensive study of the transport, specific heat, and magnetic
susceptibility on nearly stoichiometric single crystals of
Na$_{1-\delta}$FeAs. We find that Na$_{1-\delta}$FeAs single
crystal undergoes three successive phase transitions at 52, 41 and
23 K, which might correspond to structural, magnetic and
superconducting transitions, respectively. Hall coefficient
measurement indicates the development of energy gap at low
temperature in Na$_{1-\delta}$FeAs, which is similar to LaFeAsO
and SrFe$_2$As$_2$. This is consistent with the expectation of
recent density functional calculation which shows the SDW
instability for stoichiometric NaFeAs.\cite{Jishi} The present
work provides a strong evidence that a stoichiometric NaFeAs has a
magnetic ground state, similar to the parent compounds of ``1111"
and ``122".

High quality single crystals of Na$_{1-\delta}$FeAs have been
grown by the self-flux technique. The starting compositions were
selected as Na$_{1.5}$FeAs. The mixtures of Na, and FeAs were put
into an alumina crucible and sealed in Ta crucible under 2
atmosphere of argon gas. The Ta crucible was then sealed in an
evacuated quartz ampoule and heated to 1100 $^{\circ}C$ and cooled
slowly (at 5 $^{\circ}C/h$) to grow single crystals. The obtained
crystals with sizes up to 8mm$\times$5mm$\times$0.5mm have the
form of platelets with shinny surface. These crystals were
characterized by X-ray diffraction (XRD). Figure 1 shows the X-ray
diffraction pattern of Na$_{1-\delta}$FeAs with the 00$\ell$
reflections. The lattice constant c = 7.028 $\AA$ was calculated
from the higher order peaks, comparable to that of polycrystalline
sample.\cite{Parker} The elemental composition of the single
crystal was checked by Inductively Coupled Plasma (ICP) analysis.
Several crystals from the same batch were analyzed and the
deficiency of sodium was found to be less than 1$\%$; that is, the
elemental composition of the single crystal is very close to a
stoichiometric 1:1:1. The resistivity was measured by a standard
4-probe method. The DC magnetic susceptibility was measured with a
magnetic field of 0.1 T. The Hall coefficient measurement was done
using a five-probe technique. The specific heat measurement was
carried out using a thermal relaxation calorimeter. These
measurements were performed down to 2 K in a Physical Property
Measurement System(PPMS) of Quantum Design.

\begin{figure}
\includegraphics[width=8.5cm,clip]{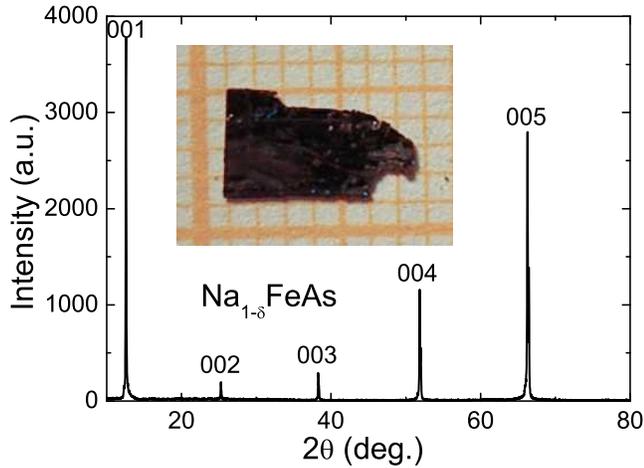}
\caption{Single crystal X-ray diffraction pattern for
Na$_{1-\delta}$FeAs. The inset shows the photograph of a
Na$_{1-\delta}$FeAs single crystal (length scale 1 mm).}
\end{figure}

Figure 2(a) shows the temperature dependence of in-plane
resistivity $\rho_{ab}$ of Na$_{1-\delta}$FeAs at zero field. At
high temperature, $\rho_{ab}$ shows a metallic behavior, whereas
$\rho_{ab}$ increases steeply below 52 K with a shoulder around 41
K, and then shows a superconducting transition near 23 K (drops to
zero resistivity at 8 K). The superconducting transition seems to
be broad, similar to that of polycrystal (not shown here). Here
the former two anomalies may correspond to the structural/magnetic
transitions. The increase of the electrical resistivity at the
phase boundary might be attributed to the opening of a gap on part
of the Fermi surfaces. This gap formation is very likely to be
induced by the occurrence of an antiferromagnetic (AF) SDW,
similar to the ``1111" compounds. One consequence of this gap is
to reduce the effective number of conduction electrons. We have
also attempted to probe the influence of magnetic fields H on
$\rho$(T) with H up to 14 T, as shown in Fig. 2(b) and (c). We
find that the magnetic phase transition temperature is insensitive
to the applied field H and $\rho$ is field-independent in the AF
state for H$\parallel$ab-plane (I$\parallel$ab). However, when H
is applied along c-axis,
[$\rho_{ab}(14T)$-$\rho_{ab}(0T)$]/$\rho_{ab}(0T)$ reaches as high
as 18$\%$ at 25 K. The large positive magnetoresistance was also
observed in ``1111" and ``122" compounds.\cite{Chen2, Dong} Note
that $\rho$ increases more rapidly below 40 K at high magnetic
fields, one might speculate that the magnetic order is formed at
this temperature, and the structural phase transition occurs at
higher temperature of 50 K in Na$_{1-\delta}$FeAs, in analogy to
``1111" compounds, where the neutron experiment has revealed that
the structural phase transition occurs prior to magnetic ordering.

\begin{figure}
\includegraphics[width=8.5cm,clip]{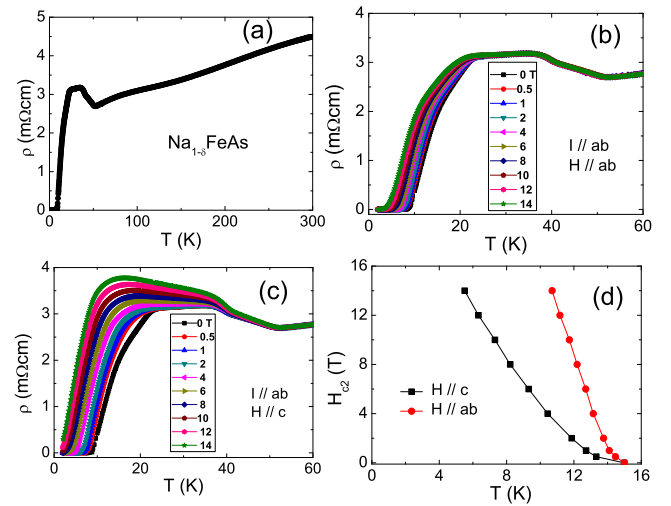}
\caption{(Color online) (a) The in-plane resistivity $\rho_{ab}$
for Na$_{1-\delta}$FeAs at zero field. Temperature dependence of
the in-plane electrical resistivity for Na$_{1-\delta}$FeAs in low
temperature region at fixed fields up to 14T for (b) H
$\parallel$ab plane and (c) H $\parallel$c-axis. (d)H$_{c2}$(T)
plot for H $\parallel$ab plane (closed square) and H
$\parallel$c-axis (closed circle), respectively.}
\end{figure}

Now let us see the magnetic field effect on superconductivity as
shown in Fig. 2(b)-(d). We can find the superconducting transition
is broadened slightly in magnetic fields up to 14 T, although it
has a broad transition at zero field. This behavior is rather
different from polycrystalline LaFeAsO where the superconducting
transition is broadened strongly in magnetic fields due to the
weak link between superconducting grains.\cite{Chen3} Figure 2(d)
shows H$_{c2}$-T$_c$ curves for both H$\parallel$ab and
H$\parallel$c, respectively, where T$_c$ is defined by a criterion
of 50$\%$ of normal state resistivity. The curves H$_{c2}$(T) are
very steep with slopes -dH$_{c2}^{ab}$ /dT $\mid _{T_c}$=3.98 T/K
for H$\parallel$ab and -dH$_{c2}^{c}$/dT$\mid_{T_c}$=2.19 T/K for
H$\parallel$c. Using the Werthamer-Helfand-Hohenberg
formula\cite{WHH} H$_{c2}$(0)=-0.69(dH$_{c2}$/dt)T$_c$ and taking
T$_c$=15 K, the upper critical fields are estimated as
H$_{c2}^{ab}$=59.7 T and H$_{c2}^{c}$=32.8 T, respectively.
Regarding the relatively low value of T$_{c}$, the upper critical
fields H$_{c2}$(0) seem to be very high. The anisotropy ratio
$\gamma$=H$_{c2}^{ab}$/H$_{c2}^{c}$$\approx$1.8 is rather small,
which is close to that of Sr$_{0.6}$K$_{0.4}$Fe$_2$As$_2$ with
$\gamma$ $\approx$ 2.0. \cite{Chen2} It is much lower than high
T$_c$ cuprates, for example $\gamma$ $\approx$ 7-10 for
YBCO.\cite{Nanda} The lower value of $\gamma$ indicates that the
inter-plane coupling in Na$_{1-\delta}$FeAs is relative strong and
the energy band with strong dispersion along z-direction may play
an important role in understanding the superconductivity of
Fe-based superconductors.

\begin{figure}
\includegraphics[width=8.5cm]{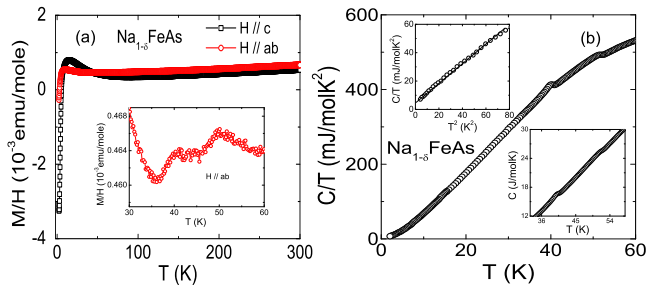}
\caption{(Color online) (a)  Magnetic susceptibility of
Na$_{1-\delta}$FeAs as a function of temperature with
H$\parallel$ab-plane and H$\parallel$c-axis, respectively. The
inset shows the enlarged part of $\chi$ between 30 and 60 K. (b)
Temperature dependence of specific heat divided by temperature C/T
for Na$_{1-\delta}$FeAs. The upper inset shows the fit of C/T vs.
T$^{2}$ in the low temperature region. The lower inset shows the
plot of C vs. T around the structural and magnetic phase
transitions.}
\end{figure}

In Fig. 3(a), we present the temperature dependence of magnetic
susceptibility $\chi$ for Na$_{1-\delta}$FeAs in a field of 0.1 T
with H$\parallel$ab-plane and H$\parallel$c-axis, respectively. At
high temperature, $\chi$ decreases monotonically with decreasing
temperature. This non-Pauli and non-Curie-Weiss-like paramagnetic
behavior, is consistent with those observed in ``1111" and ``122"
parent compounds above T$_{SDW}$.\cite{Kamihara08, Chen2, GMZhang}
At low temperature, $\chi$ tends to show a small upturn, and
followed by a clear diamagnetic drop corresponding to
superconductivity at around 10 K, which corresponds to the zero
resistivity temperature. To show the low temperature part clearly,
we have plotted in the inset of Fig. 3(a) the susceptibility in
the temperature interval from 30 to 60 K. Two humps found in
$\chi$ with magnetic field H parallel to ab-plane at 40 and 50 K
give that the magnetic easy axis of this compound is along
ab-plane (There is no detectable anomaly observed in $\chi$ with
H$\parallel$c). That is to say, the Fe spins in NaFeAs order
antiferromagnetically with spin direction parallel to ab-plane,
similar to that of SrFe$_2$As$_2$ which has been confirmed by
neutron scattering experiment.\cite{ZhaoJ}

\begin{figure}
\includegraphics[width=8.5cm, clip]{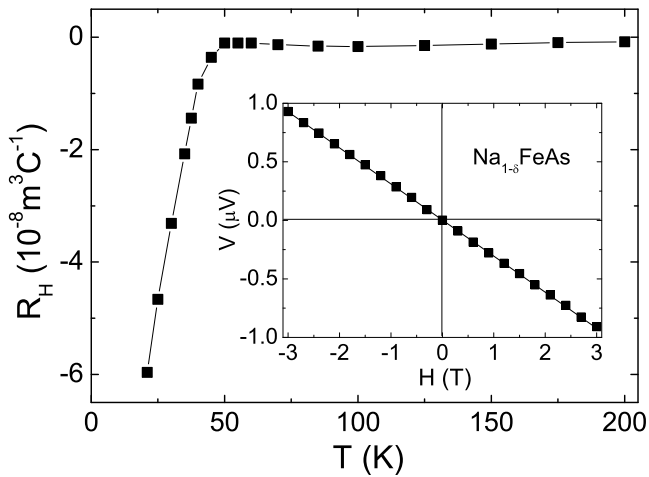}
\caption { Temperature dependence of Hall coefficient for
Na$_{1-\delta}$FeAs. The inset shows the verification of the Hall
voltage driven by magnetic field where a linear dependence of the
transverse voltage on the applied magnetic field is observed up to
3 T at 40 K.}
\end{figure}

To get more information about the structural/magnetic phase
transition, we performed specific heat measurements for
Na$_{1-\delta}$FeAs. Figure 3(b) shows the temperature dependence
of C/T from 2 to 60 K. Two successive jumps in C/T at
T$_{1}$$\sim$41 K and T$_{2}$$\sim$52 K show the bulk nature of
phase transitions which have been observed in susceptibility and
resistivity data as well. However, the anomalies around T$_{1}$
and T$_{2}$ are very broad and small, which would be
characteristic of a second-order transition. Very similar
properties have been reported on ``1111" compounds, such as
LaFeAsO, where two subsequent jumps at 155 K and 143 K were
observed in specific heat data corresponding to the structural and
SDW transitions, respectively.\cite{McGuire} While the structural
transition and SDW transition occur at same temperature in
SrFe$_2$As$_2$. The specific heat shows a very sharp peak which is
the characteristic feature of first-order phase
transitions.\cite{Chen2} The conversion of first order transition
to second order transition has also been observed in those of
slightly underdoped samples of BaFe$_{2-x}$Co$_x$As$_2$ (x$<$0.2),
in which the single structural/magnetic phase transition splits
into two distinct phase transitions (The first-order phase
transition starts rounding off and becomes smaller as more Co ions
are added in the pure system).\cite{Chu} The onset of a second
order phase transition is probably related to a change of symmetry
in the ground state.\cite{HuJ, Xu} Theoretical studies also
suggest that whether the structural and magnetic transitions occur
simultaneously or separately depends on the interlayer
coupling.\cite{HuJ, Xu} The separation of structural distortion
and the magnetic transition in the present compound indicates that
the interlayer coupling is closer to that of ``1111" compounds,
but much weaker than ``122" compounds.

At low temperature, there is no detectable anomaly observed around
T$_c$, indicating a small superconducting volume fraction; the fit
of C/T vs.T$^2$ yields the electronic coefficient $\gamma$
$\approx$ 5 mJ/mol.K$^2$, as shown in the upper inset of Fig.
3(b). Here the electronic coefficient for Na$_{1-\delta}$FeAs is
close to the values for LaFeAsO and SrFe$_2$As$_2$ (which are
significantly smaller than the theoretical calculations due to the
SDW partial gap).\cite{Dong, Chen2}

The Hall coefficient R$_H$ as a function of temperature between 20
and 200 K for Na$_{1-\delta}$FeAs is shown in Fig.4. The Hall
coefficient is negative at all temperatures, indicating conduction
carriers are dominated by electrons. Above 50 K, the Hall
coefficient is nearly temperature independent; the carrier density
is estimated being n $\approx$ 9$\times$10$^{21}$ cm$^{-3}$ at 200
K if the one-band model is simply adopted. It is comparable to
that of SrFe$_{2}$As$_{2}$ with n $\approx$ 1.5$\times$10$^{22}$
cm$^{-3}$ at 300 K obtained by the same method. The large carrier
number for Na$_{1-\delta}$FeAs indicates that it is a good metal.
Noted that the band calculation on its sister compound of LiFeAs
revealed its semimetallic behavior. As mentioned above, there is a
possibility that Na deficiencies serve as a source of effective
hole carriers. So it is an extremely rare case the measurement of
a negative Hall coefficient in nominally hole-doped Fe-based
superconductors. However, this behavior could be qualitatively
understood by a simple two-band approximation. In a scenario where
the Fermi surface contains both electron and hole pockets, the
sign of R$_H$ depends on the relative magnitude of the respective
densities, n$_e$ and n$_h$, and mobilities, $\mu$$_e$ and
$\mu$$_h$ ($\mu$=e$\tau$/m*, where e is the electron charge,
1/$\tau$ is the scattering rate and m* is the effective mass). And
therefore, in the ``hole"-doped sample, the negative R$_H$ implies
that $\mu$$_e$ is much larger than $\mu$$_h$. Below 50 K, R$_H$
drops dramatically to a very large negative value. The absolute
value of R$_H$ at 23 K is about 60 times larger than that at 60 K.
The huge increase of the R$_H$ value is seen in the undoped
compounds of ``1111" and ``122", which is naturally explained by
the gapping of the Fermi surface which removes a large part of
free carriers. Evidence for the gap is also observed by optical
spectroscopy measurement.\cite{Hu} The band calculation shows that
a hole pocket is lying around the center of Brillouin zone (BZ)
and the electron pockets are lying at the BZ corners. Therefore,
the low temperature behavior of the Hall effect for
Na$_{1-\delta}$FeAs may be understood with the following scenario:
with temperature down to below T$_{SDW}$, the hole pocket is
almost fully gapped while the electron pockets are partially
gapped. The negative Hall coefficient R$_{H}$ reflects mainly the
the un-gapped electron density at the BZ corner.

In summary, we have succeeded in growing the high quality single
crystal of nearly stoichiometric Na$_{1-\delta}$FeAs and studied
the electronic properties by measurements of electrical
resistivity, heat capacity, and Hall effect. This compound is
found to undergo the structural, magnetic and superconducting
transitions at low temperatures. The negative Hall coefficient
R$_{H}$ suggests that the electron type charge carriers dominate
the conduction in this material. Although the present experiments
cannot completely rule out a tiny amount of Na deficiencies, our
present data provides the first direct experimental evidence
confirming the existence of SDW instability in NaFeAs, which are
consistent with the theoretical predication. We believe that our
results are important to understand the mechanism of
superconductivity and underline further the importance of magnetic
fluctuations for the superconductivity pairing observed in
Fe-based superconductors.

This work is supported by the NSFC, CAS, and the 973 project of
the MOST of China.


\begin{thebibliography}{20}

\bibitem{Kamihara08} Y. Kamihara, T. Watanabe, M. Hirano, and H.
Hosono, J. Am. Chem. Soc. \textbf{130}, 3296 (2008).

\bibitem{Chen1} G.F. Chen, Z. Li, D. Wu, G. Li, W. Z. Hu, J. Dong,
P. Zheng, J.L. Luo, and N.L. Wang,  Phys. Rev. Lett. \textbf{100},
247002 (2008).

\bibitem{ChenXH}X.H. Chen, T. Wu, G. Wu, R.H. Liu, H. Chen, and D.F. Fang,
Nature \textbf{453}, 761 (2008).

\bibitem{Ren1} Z.A. Ren, J. Yang, W. Lu, W.
Yi, X.L. Shen, Z.C. Li, G.C. Che, X.L. Dong, L.L. Sun, F. Zhou,
and Z.X. Zhao, Europhys. Lett. \textbf{82}, 57002 (2008).

\bibitem{Wang}  C. Wang, L.J. Li, S. Chi, Z.W. Zhu, Z. Ren, Y.K. Li,
Y.T. Wang, X. Lin, Y.K. Luo, S. Jiang, X.F. Xu, G.H. Cao, and Z.A.
Xu, Europhys. Lett. \textbf{83}, 67006 (2008).

\bibitem{Rotter2} M. Rotter, M. Tegel, D. Johrendt,
Phys. Rev. Lett. \textbf{101}, 107006 (2008).

\bibitem{Chen} G.F. Chen, Z. Li, G. Li, W.Z. Hu, J. Dong, X.D. Zhang, P.
Zheng, N.L. Wang, and J.L. Luo, Chinese Phys. Lett. \textbf{25},
3403 (2008).

\bibitem{Canfield} M.S. Torikachvili, S.L. Bud'ko, N. Ni, P.C.
Canfield, Phys. Rev. Lett., \textbf{101}, 057006 (2008).

\bibitem{Dong} J. Dong, H.J. Zhang, G. Xu, Z. Li, G. Li, W.Z. Hu, D. Wu,
G.F. Chen, X. Dai, J.L. Luo, Z. Fang, and N.L. Wang, Europhys.
Lett. \textbf{83}, 27006 (2008).

\bibitem{Jin1} X.C. Wang, Q.Q. Liu, Y.X. Lv, W.B. Gao, L.X. Yang, R.C. Yu, F.Y.
Li, and C.Q. Jin,  Solid State Comm. \textbf{148} 538 (2008).

\bibitem{Zhu1} J.H. Tapp, Z. Tang, B. Lv, K. Sasmal, B. Lorenz,
Paul C. W. Chu, and A.M. Guloy, Phys. Rev. B \textbf{78} 060505(R)
(2008).

\bibitem{Zhu2} C.W.Chu, F. Chen, M. Gooch, A. M. Guloy, B. Lorenz, B.
Lv, K. Sasmal, Z.J. Tang, J.H. Tapp, and Y.Y. Xue,
arXiv:0902.0806.

\bibitem{Parker} D.R. Parker, M.J. Pitcher, P.J. Baker, I. Franke, T. Lancaster,
S.J. Blundell, and S.J. Clarke,  Chem. Commun., 2009, 2189-2191.


\bibitem{Jishi}  R.A. Jishi, and H.M. Alyahyaei, arXiv:0812.1215.

\bibitem{Singh}  D.J. Singh, Phys. Rev. B \textbf{78} 094511
(2008).

\bibitem{Nekra} I.A. Nekrasov, Z.V. Pchelkina, and M.V. Sadovskii, JETP Lett.
\textbf{88} 543 (2008).

\bibitem{Chen2} G.F. Chen, Z. Li, J. Dong, G. Li, W.Z. Hu, X.D. Zhang, X.H. Song,
P. Zheng, N.L. Wang, and J.L. Luo,  Phys. Rev. B \textbf{78},
224512 (2008).

\bibitem{Dai} Clarina de la Cruz, Q. Huang, J.W. Lynn, J.Y. Li, W. Ratcliff
II, J.L. Zarestky, H.A. Mook, G.F. Chen, J.L. Luo, N.L. Wang, P.C.
Dai, Nature \textbf{453} 899 (2008).


\bibitem{Chen3} G. F. Chen, Z. Li, G. Li, J. Zhou, D. Wu, J. Dong, W. Z. Hu, P.
Zheng, Z. J. Chen, H. Q. Yuan, J. Singleton, J. L. Luo, and N. L.
Wang, Phys. Rev. Lett. \textbf{101}, 057007 (2008).

\bibitem{WHH} N.R. Werthamer, E. Helfand and P.C. Hohenberg,
Phys. Rev. \textbf{147}, 295(1966).

\bibitem{Nanda} K.K. Nanda, Physica C \textbf{265}, 26 (1996).

\bibitem{GMZhang} G.M. Zhang, Y.H. Su, Z.Y. Lu, Z.Y. Weng, D.H. Lee, and T.
Xiang, arXiv:0809.3874 and references therein.

\bibitem{ZhaoJ} J. Zhao, D.X. Yao, S. Li, T. Hong, Y. Chen, S. Chang,
W. Ratcliff, J.W. Lynn, H.A. Mook, G.F. Chen, J.L. Luo, N.L. Wang,
E.W. Carlson, J.P. Hu, and P.C. Dai, Phys. Rev. Lett.
\textbf{101}, 167203 (2008).

\bibitem{McGuire} M.A. McGuire, A.D. Christianson, A.S. Sefat, B.C. Sales et al.,
Phys. Rev. B \textbf{78} 094517 (2008).

\bibitem{Chu} J.H. Chu, J.G. Analytis, C. Kucharczyk, and I.R.
Fisher, Phys. Rev. B \textbf{79} 014506 (2009).

\bibitem{HuJ} C. Fang, H. Yao, W.F. Tsai, J.P. Hu and S.A.
Kivelson, Phys. Rev. B \textbf{77} 224509(2008).

\bibitem{Xu} Y. Qi and C.K. Xu,  arXiv:0812.0016.

\bibitem{Hu} W.Z. Hu, et al. unpublished


\end{thebibliography}
\end{document}